\newcommand{\eps}{\epsilon}
\newcommand{\GeV}{\, \rm{GeV}}
\newcommand{\eV}{\, \rm{eV}}

\documentclass{PoS}

\title{Flavored Axions}

\ShortTitle{Flavored Axions}

\author{Robert Ziegler 
\\

        CERN, Theoretical Physics Department, 1 Esplanade des Particules, Geneva 23, Switzerland \\
        E-mail: \email{robert.ziegler@cern.ch}}

\abstract{Precision flavor experiments can look for the QCD axion complementarily to usual searches with axion helio- and haloscopes, allowing to test PQ breaking scales as high as $10^{12} \GeV$. Such searches are sensitive to flavor-violating axion couplings, which are generic and potentially sizable whenever SM fermions carry flavor non-universal PQ charges. A particularly predictive scenario is obtained when PQ is identified with the simplest FN flavor symmetry, so that all flavor-violating axion couplings are related to Yukawa hierarchies, up to ${\cal O}(1)$ coefficients.  
}

\FullConference{Corfu Summer Institute 2018 "School and Workshops on Elementary Particle Physics and Gravity"\\
		(CORFU2018)\\
		31 August - 28 September, 2018\\
		Corfu, Greece}

\begin{document}


\section{Introduction}
One of the best motivated particle beyond the SM is arguably the QCD axion. Originally proposed as a solution to the strong CP Problem~\cite{PQ1,PQ2,WW1,WW2}, it was soon realized that the axion can also serve as a viable Dark Matter candidate in vast parts of the parameter space~\cite{AxionDM1, AxionDM2, AxionDM3}. In the past years axion searches have received renewed interest, and many new ideas have been pushed forward to look for the axion with small-scale experiments (see Ref.~\cite{ExpReviewGraham} for a review). Besides the imperative axion couplings to gluons, most searches employ the axion couplings to photons, which are expected to be sizable in the majority of axion models~\cite{Luca1, Luca2}. Although more sensitive to the underlying UV axion model, also \emph{flavor-violating axion couplings}  allow to look for the QCD axion in rare decays with high-precision flavor experiments. In the following we discuss various aspects of such flavored axions; we begin  with a brief review of axion phenomenology and the present experimental constraints, before discussing the theoretical origin of flavor-violating axion couplings. We finally review a particularly predictive flavored axion model where the Peccei-Quinn (PQ) symmetry  is identified with the simplest flavor symmetry addressing Yukawa hierarchies.

\section{Axion Couplings}
At energies much below the PQ breaking scale, the relevant effective axion Lagrangian reads
\begin{equation}
{\cal L}  =  \frac{a}{f_a} \frac{\alpha_s}{8 \pi}  G \tilde{G} +  \frac{E}{N}  \frac{a}{f_a} \frac{\alpha_{\rm em}}{8 \pi} F \tilde{F} + \frac{\partial_\mu a}{2 f_a} \overline{f}_i \gamma^\mu \left( C^V_{ij} + C^A_{ij} \gamma_5 \right) f_j \, ,
\label{La}
\end{equation}
where $f_i$ runs over SM quark and lepton generations and $E/N$ is the ratio between the electromagnetic and color anomaly coefficients. In complete generality, the axion couplings to fermions are parameterized by two hermitian $3\times3$ matrices $C^V$ and $C^A$ in each fermion sector. Note that flavor-diagonal  vector couplings are unphysical, because they can be absorbed by non-anomalous field redefinitions. 

The first term in Eq.~(\ref{La}) defines the axion decay constant $f_a$, and gives the only contribution to the axion potential, which can be conveniently calculated using chiral perturbation theory. This potential has a trivial minimum, thus dynamically setting the QCD $\theta$-term to zero, which explains the absence of CP violation in strong interactions. The same potential generates an axion mass whose parametric size is set by $m_a \propto m_\pi f_\pi/f_a$, and including higher-order corrections~\cite{Villadoro} given by
\begin{equation}
m_a  = 5.7 \, {\rm \mu eV} \left( \frac{10^{12} \GeV}{f_a} \right) \, .
\end{equation}
The second term in Eq.~(\ref{La}) gives rise to axion couplings to photons, which at energies much below the QCD scale are given by~\cite{Villadoro}
\begin{equation}
{\cal L}  \supset  C_\gamma \frac{a}{f_a} \frac{\alpha_{\rm em}}{8 \pi} F_{\mu \nu} \tilde{F}^{\mu \nu}  \, , \qquad \qquad C_\gamma = \left| E/N - 1.92(4) \right| \, .
\label{Lagamma}
\end{equation}
These couplings are constrained by astrophysics~\cite{RaffeltAstro}, in particular from  the evolution of HB stars in globular clusters~\cite{HBbounds}. This bound, which is of the same order as the constraint from the CAST experiment~\cite{CAST}, translates into a lower bound on  $C_\gamma/f_a$ or equivalently an upper bound on $C_\gamma m_a$
\begin{equation}
m_a < \frac{0.3}{C_\gamma} \eV \, .
\end{equation}
The next generation  of  axion  helioscopes  such as IAXO~\cite{IAXO1, IAXO2, IAXO3} will be able to improve this bound by about an order of magnitude. 

Turning to the third term in Eq.~(\ref{La}), it is again astrophysics that provides the strongest constraints on (flavor-diagonal) axion couplings to ordinary matter, i.e. nucleons and electrons. 
From the shape of the White Dwarf luminosity function~\cite{WDbound} one obtains a bound on the axion coupling to electrons $C_e \equiv |C_{ee}^A|$
\begin{equation}
m_a < \frac{3 \cdot 10^{-3}}{C_e} \eV \, ,
\end{equation}
while the burst duration of the SN 1987A neutrino signal provides a constraint on the axion coupling to nucleons~\cite{SNbound}, defined as 
\begin{equation}
{\cal L}  = \frac{\partial_\mu a}{2 f_a} \left[ C_p \overline{p} \gamma^\mu \gamma_5 p + C_n \overline{n} \gamma^\mu \gamma_5 n \right] \, .  
\end{equation}
The proton and neutron couplings $C_{p,n}$ in turn are given by the (flavor-diagonal) axial vector couplings to quarks $C^A_{qq}$, apart from a model-independent contribution due to axion-gluon couplings~\cite{Villadoro}. Using the bound from Ref.~\cite{CoolingAnom3}, the average axion couplings to nucleons defined as $C_N \equiv \sqrt{C_p^2 + C_n^2}$ is constrained at a level comparable to the bound on electron couplings
\begin{equation}
m_a < \frac{4 \cdot 10^{-3}}{C_N} \eV \, .
\end{equation}
Finally we discuss the constraints on flavor off-diagonal axion couplings, which arise from flavor-violating decays with invisible and practically massless final state axions. This signature is very similar to very rare meson decays in the SM like $K \to \pi \nu \overline{\nu}$, which are strongly constrained by experiments. Using the latest bound from E787/E949 on $K \to \pi a$~\cite{Kpia}, one obtains for $s-d$ transitions  
\begin{equation}
m_a < \frac{2 \cdot 10^{-5}}{|C^V_{sd}|} \eV \, .
\end{equation}
Interestingly this (already stringent) bound could be further improved in the near future with the NA62 experiment by almost an order of magnitude~\cite{NA621, NA622}, which allows to test axion decay constants up to $10^{12} \GeV$. Similarly, $b-s$ transitions are constrained from $B \to K a$ searches at CLEO~\cite{BKa}, giving 
\begin{equation}
m_a < \frac{9 \cdot 10^{-2}}{|C^V_{bs}|} \eV \, .
\end{equation}
Also this bound can be further improved by the BELLE II experiment in the near future, presumably by at least an order of magnitude. Turning to the charged lepton sector, the experimental situation becomes more challenging, as the main decay channels in the SM are similar to the signal. Nevertheless constraints on e.g. $\mu-e$ transitions have been obtained in the late 80's with the Crystal Box detector~\cite{mueaga1, mueaga2}, which allows to put bounds on the averaged couplings $C_{\mu e} \equiv \sqrt{|C_{\mu e}^V|^2 + |C_{\mu e}^A|^2 }$
  \begin{equation}
m_a < \frac{3 \cdot 10^{-3}}{C_{\mu e}} \eV \, .
\end{equation}
These bounds are likely to be further improved by the MEG II~\cite{MEG2} and/or Mu3e experiments~\cite{Mu3e}.

\begin{figure}
 \centering
     \includegraphics[width=.6\textwidth]{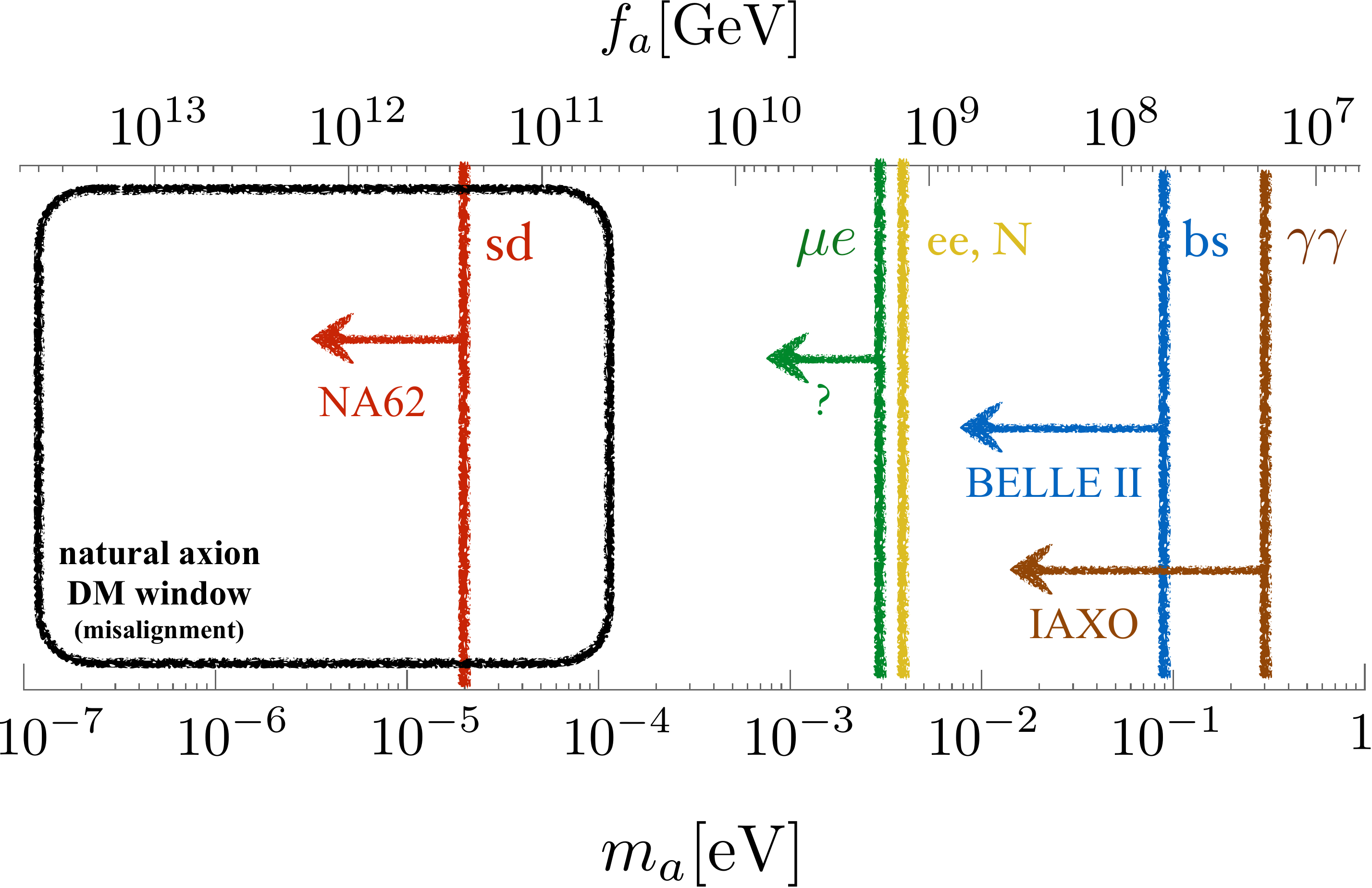}
     \caption{Sketch of present and future constraints on axion couplings for $C_i = 1$, see text for details.  }
     \label{fig1}
     \end{figure}

A summary plot of the most relevant constraints\footnote{Constraints on the remaining flavor-violating axion couplings to quarks and leptons can be found in Refs.~\cite{Murayama, King}.} discussed so far is shown in Fig.~\ref{fig1}, which shows the upper bound on the axion mass from various processes by setting the respective dimensionless couplings $C_i = \{ C_\gamma, C_e, C_N, C^V_{sd}, C^V_{bs} \}$   to 1. Also shown is the region where the axion can naturally account for the present Dark Matter abundance through the misalignment mechanism~\cite{AxionDM1, AxionDM2, AxionDM3}. 

From Fig.~\ref{fig1} it is clear that rare flavor-violating decays are very important to constrain flavor-violating axion couplings to matter, and can compete with the stringent constraints   
 on flavor-diagonal couplings from astrophysics. In the case of $s-d$ transitions, the present bounds on $f_a$ from $K \to \pi + a$ decays are actually about two orders of magnitude stronger, for equal sizes of the respective dimensionless couplings. Most interestingly, these bounds are expected to be improved in the very near future by various experiments such as NA62 and Belle II. Therefore precision flavor experiments provide the exciting possibility to look for  the QCD axion in a way that is complementary to the usual searches with helio- and haloscopes.   
\section{Flavored Axions}
In this section we investigate the expected size of flavour-violating axion couplings in UV axion models. In general these couplings arise from the PQ current, which has to be rotated to the fermion mass basis. In this basis, given by unitary rotations $V_{f}$ defined by $V_{f_L}^\dagger M_f V_{f_R}= M_f^{\rm diag}$, one obtains
\begin{equation}
C_{f_i f_j}^{V,A}  = \frac{1}{2N} \left( V_{f_R}^\dagger X_{f_R} V_{f_R} \pm V_{f_L}^\dagger X_{f_L} V_{f_L} \right)_{ij}  \, ,
\label{Cdef}
\end{equation}
where $X_{f_L, f_R}$ denote the (flavor-diagonal) PQ charge matrices of left- and right-handed fermions. Therefore flavor-violating couplings are present whenever SM fermions carry PQ charges that represent a new source of flavor violation beyond SM Yukawas, i.e. \emph{whenever PQ charges do not commute with Yukawa matrices}. In this case the off-diagonal couplings depend on the unitary rotations that connect interaction and mass basis, and thus can be quantitatively predicted only in a theory of flavor. 

 The minimal scenario is to completely disentangle the solution to the strong CP problem from the origin of SM flavor puzzle, i.e. axion and flavor physics, as in common axion benchmark models. Indeed in KSVZ models~\cite{KSVZ1,KSVZ2} SM fermions do not carry PQ charges, and thus do not couple to the axion at all\footnote{Still axion couplings to nucleons are induced as a result of the axion couplings to gluons.}. In the simplest DFSZ models~\cite{DFSZ1, DFSZ2} SM fermions are given flavor-universal PQ charges, which also implies that flavor-violating couplings vanish at tree-level.   
 
The simplest example of scenarios with flavor-violating axion couplings are DFSZ models with non-universal PQ charges (which of course provide an equally good solution to the strong CP problem as long as PQ is anomalous under QCD). In such cases the flavor-violating axion couplings depend on the unitary rotations, but can be effectively parametrized by a couple of free parameters under certain assumptions, e.g. if PQ allows all Yukawa couplings and only two Higgs doublets are present. Such scenarios can be motivated by other  features, for example the possibility to suppress the axion couplings to nucleons and/or electrons~\cite{Astrophobic}. 

More predictive are scenarios where PQ is also (partially) responsible for explaining the peculiar pattern of SM Yukawas, which has been proposed already long time ago by F.~Wilczek~\cite{Wilczek}. A particular simple realization\footnote{For other possibilities in the context of larger flavor symmetry groups see e.g. Refs.~\cite{Celis1, Matthias, Carone}, } is based on the identification of the PQ symmetry with the smallest flavor symmetry able to address Yukawa hierarchies, i.e. a horizontal $U(1)$ symmetry à la Froggatt-Nielsen (FN)~\cite{Japs, Axiflavon}.
In the next section we briefly summarize the scenario in Ref.~\cite{Axiflavon}. 

\section{PQ=FN: The Axiflavon}
As in usual FN models~\cite{FN, SeibergNir1} we assume that the hierarchies of the Yukawa couplings are due to a global horizontal symmetry $U(1)_H$, under which SM Weyl fermions carry positive, flavor-dependent charges $[q]_i, [u]_i, [d]_i, [l]_i, [e]_i$, respectively, while the Higgs is neutral. The $U(1)_H$ symmetry is spontaneously broken at high scales by the vev $V_\Phi$ of a complex scalar field $\Phi$ with $U(1)_H$ charge of $-1$. SM Yukawas arise then from higher-dimensional operators involving appropriate powers of $\Phi$ to be invariant under $U(1)_H$, suppressed by the UV cutoff scale $\Lambda$. After plugging in the vev of $\Phi = V_\Phi/\sqrt{2}$, this gives rise to SM Yukawa couplings 
\begin{equation}
y^{u,d,e}_{ij}  = a^{u,d,e}_{ij} \eps^{[L]_i + [R]_j} \,,
\label{yuks}
\end{equation}
where $[L]_i = [q]_i, [R]_i = [u]_i, [d]_i$ in the quark sectors, $[L]_i = [l]_i, [R]_i = [e]_i$ in the charged lepton sector and we have defined the small parameter $\eps \equiv V_\Phi/(\sqrt 2 \Lambda)$. Here $a^{u,d,e}_{ij}$ are unknown Wilson coefficients, of the effective operators, assumed to be ${\cal O}(1)$. While to some extent they can be determined together with the fermion charges by a numerical fit to fermion masses and mixings, here we focus on analytical results that aim to incorporate the uncertainties present in such a fit. 

In this scenario, upon the identification of PQ with $U(1)_H$,  the Goldstone boson contained in $\Phi$ plays the role of the QCD axion, and its couplings to gluons, photons and fermion are determined by the horizontal fermion charges. Although the precise value of these charges depend on the fermion mass fit, one can find a pretty narrow range for the ration of electromagnetic and color anomaly coefficient, given by 
 \begin{equation}
\frac{E}{N}  \in \left[ 2.4 , 3.0 \right] \, .
\label{ENpred}
\end{equation}
This surprisingly narrow window can be obtained by calculating the determinants of the fermion mass matrices, and expressing them in terms of the anomaly coefficients. This leads to 
\begin{equation}
\frac{E}{N} = \frac{8}{3} - 2 \frac{\log \frac{{\rm det} \,  m_d}{{\rm det} \,  m_e} - \log \alpha_{de}}{\log \frac{{\rm det} \,  m_u {\rm det} \,  m_d}{v^6} - \log \alpha_{ud}} \, ,
\label{ENprediction}
\end{equation}
where $\alpha_{ud} = {\rm det} \,  a_u {\rm det} \, a_d $ and $\alpha_{de} = {\rm det} \,  a_d/{\rm det} \,  a_e$ contain the ${\cal O}(1)$ uncertainties from the coefficients in Eq.~(\ref{yuks}). Independently of the precise values of these parameters, it is clear that the second term on the right-hand side of Eq.~(\ref{ENprediction}) is strongly suppressed by the large denominator ($\log {\rm det} \,  m_u {\rm det} \,  m_d/v^6 \approx -44 $), so that $E/N$ is expected to be close to $8/3$.

Axion couplings to fermions arise from Eq.~(\ref{Cdef}), with unitary rotations that are themselves determined (at least parametrically) by $U(1)_H$ charges according to
\begin{equation}
(V_{f_L})_{ij} \sim  \eps^{|[L]_i - [L]_j|} \, ,  \qquad \qquad (V_{f_R})_{ij} \sim  \eps^{|[R]_i - [R]_j|} \, .
\end{equation}
Therefore all flavor-diagonal couplings are expected to be ${\cal O}(1)$, while off-diagonal couplings are suppressed by small rotation angles. Using typical values of $U(1)_H$ charges needed to give a good fit to quark masses and mixings, it is clear that flavor-violating $s-d$ couplings are just suppressed by a Cabibbo angle and therefore sizable
\begin{equation}
C^V_{sd} \sim \lambda \approx 0.2.
\end{equation}
This implies an upper bound on the axion mass of about $m_a \lesssim 10^{-4} \eV$, which is just  
at the edge of the natural axion DM window (see Fig.~(\ref{fig1})). Thus the ``axiflavon" in this scenario will be tested in the near future complementarily by precision flavor physics at NA62 and axion haloscopes with the ADMX upgrade~\cite{ADMXfuture}. 
\section{Conclusions}
To summarize, we have argued that precision flavor experiments allow to look for the QCD axion complementarily to the usual searches with axion helio- and haloscopes. The strongest sensitivity is obtained for flavor-violating $s-d$ transitions, where NA62 is expected to test PQ breaking scales as high as $10^{12} \GeV$ by looking for $K \to \pi a$ decays. Such decays are sensitive to flavor-violating axion couplings, which are generic and potentially sizable whenever SM fermions carry flavor non-universal PQ charges. In this case the axion couplings depend on the misalignment of PQ charges and Yukawa couplings, i.e. a theory of flavor is need in order to make quantitative predictions for off-diagonal couplings. A particularly predictive scenario is obtained when PQ is identified with the simplest FN flavor symmetry, in which case all flavor-violating axion couplings are directly related to Yukawa hierarchies, up to ${\cal O}(1)$ coefficients.  

\newpage 
\bibliographystyle{JHEP}
\bibliography{Mybib}


\end{document}